\documentclass[preprint,aps,12pt,showpacs,nofootinbib,tightenlines]{revtex4}
\usepackage{mathrsfs}
\usepackage{amsmath}
\usepackage{amssymb}
\usepackage{epsfig}
\usepackage{graphicx}
\newcommand{\psl}{ p \hspace{-1.5truemm}/ }

\newcommand{\esl}{ \epsilon \hspace{-1.5truemm}/ }

\def\be{\begin{eqnarray}}
\def\en{\end{eqnarray}}
\def\non{\nonumber\\}

\def\prd{{Phys. Rev. D}~}
\def\prl{{ Phys. Rev. Lett.}~}
\def\plb{{ Phys. Lett. B}~}

\def\epjc{{ Eur. Phys. J. C}~}

\begin{document}
\title{Insights into the nature of the $X(3872)$ through B meson decays}
\author{Zhi-Qing Zhang
\footnote{Electronic address: zhangzhiqing@haut.edu.cn}, Zhi-Lin Guan, Yan-Chao Zhao, Zi-Yu Zhang, Zhi-Jie Sun, Na Wang, Xiao-Dong Ren} 
\affiliation{\it \small  Department of Physics, Henan University of
Technology, Zhengzhou, Henan 450052, China } 
\date{\today}
\begin{abstract}
We study the decays $B_{c,u,d}\to X(3872)P$ in the perturbative QCD (PQCD) approach, where the puzzling resonance $X(3872)$
is involved and $P$ represents a light pseudoscalar meson $K$ and $\pi$. Assuming the $X(3872)$ as a $1^{++}$ charmonium state,
we find the following results: (a) The branching ratios for
the decays $B^+_c\to X(3872)\pi^+$ and $B^+_c\to X(3872) K^+$ agree with the results predicted by the covariant
light-front approach  within errors, but are larger than those given by the generalized factorization approach; (b) The branching ratio for the
decay $B^+\to X(3872)K^+$ is predicted as $(3.8^{+1.1}_{-1.0})\times10^{-4}$, which is smaller than the previous PQCD calculation result,
but still slightly larger than the upper limits set by Belle and BaBar. So we suggest that the decays $B^{0,+}\to X(3872)K^{0,+}$
should be precisely measured by the running LHCb and Belle II experiments, which is very helpful to probe the inner structure of the $X(3872)$;
(c) Compared with the decays $B_{u,d}\to X(3872)K$,  the decays $B_{u,d}\to X(3872)\pi$ have much smaller branching ratios, which drop to
as low as $10^{-6}$; (d) The direct CP violations for these considered decays are very small, only $10^{-3}\sim 10^{-2}$, because the
penguin contributions are loop suppressed compared with the tree contributions. Testing the results for the branching ratios and the CP violations
including the implicit $SU(3)$ and isospin symmetries in these decays by experiments is helpful to probe the nature of the $X(3872)$.
\end{abstract}

\vspace{1cm}

\maketitle


\section{Introduction}\label{intro}
Since $X(3872)$ was observed in the exclusive decay $B^{\pm}\to K^{\pm}\pi^+\pi^-J/\Psi$ by Belle \cite{choi},
this meson has attracted a great deal of interest up to now. Though the $X(3872)$ has been confirmed by many experimental collaborations, such as
CDF \cite{cdf}, D0 \cite{d0}, Babar \cite{babar0} and LHCb \cite{lhcb}, with quantum numbers $J^{PC}=1^{++}$ and isospin $I=0$, there are
still many uncertainties. Since the mass of the $X(3872)$ is very close to the $D^0\bar D^{*0}$ threshold, some authors interpret it
as a loosely bound molecular state \cite{close,wong,braaten0,swanson,zhusl}, where the building blocks are hadrons and they interact with each other by exchanging
color neutral forces \cite{zhenghq}. Others regard the $X(3872)$
as a compact tetraquark state \cite{chiu,barnea,maiani1,hogaasen}, where the building blocks are (anti-)quarks and they interact with each other by exchanging gluons.
Certainly, there is also explanation of the $X(3872)$ as a hybrid charmonium state with constituents $c\bar c g$ \cite{close2,li}, etc.
Though there are many different exotic hadron state interpretations about the $X(3872)$, the first raidal excitation of $1P$ charmonium
state $\chi_{c1}(1P)$ as the most natural assignment has not been ruled out \cite{barnes,eichten,quigg}. It is noticed that the $X(3872)$ is renamed as $\chi_{c1}(3872)$ by the Particle Data Group
(PDG) \cite{pdg20}.

In order to investigate the inner structure of the $X(3872)$, a large amount of theoretical studies on the
productions and decays of the $X(3872)$ were performed \cite{hxchen,meng,wuq,math}. In Ref. \cite{math}, the authors calculated
$\Gamma(X(3872)\to J/\psi \pi^+\pi^-)$
by using the QCD sum rules and concluded that the X(3872) is approximately $97\%$ a charmonium sate with a tiny molecular component. Many B meson decays with the $X(3872)$ involved in the
final state have been studied by using different approaches\cite{mengc0,mengc,WSL,hsiao,braaten,braaten2,liux}. In Ref. \cite{mengc}, the authors studied the decays $B\to \chi_{c1}(1P,2P)K$ in QCD
factorization, and they argued that the $X(3872)$ has a dominant $c\bar c$ component but mixes with $D^0\bar D^{*0}+D^{*0}\bar D^0$ continuum component.
The decays $B_c\to X(3872)\pi(K)$ were studied both in the covariant light-front approach \cite{WSL} and the generalized factorization
approach \cite{hsiao}, respectively. In the former the $X(3872)$ was identified as a $1^{++}$
charmonium sate, while a tetraquark state was assumed in the latter. One may expect that the results for the same decays under the different
structure hypothesis for the $X(3872)$ should be different. The decay $B\to X(3872)K$ was also received much attention by many authors. In
Ref.\cite{braaten,braaten2}, the authors assumed the $X(3872)$ as a loosely bound S-wave molecular sate of $D^0\bar D^{*0}(D^{*0}\bar D^0)$
and estimated $Br(B^+\to X(3872)K^+)=(0.07\sim1)\times10^{-4}$. Furthermore, they considered  that $Br(B^0\to X(3872)K^0)$ is
suppressed more than an order of magnitude compared with $Br(B^+\to X(3872)K^+)$. That is to say, if large isospin symmetry between
the decays $B^+\to X(3872)K^+$ and $B^0\to X(3872)K^0$ is observed, any charmonium interpretation for the $X(3872)$ would be disfavored.
Two years later, the branching ratio of the decay $B^+\to X(3872)K^+$ was calculated by using the PQCD approach under assuming
the $X(3872)$ as a regular $c\bar c$ charmonium state in Ref. \cite{liux}, where a large value for the branching ratio was obtained
$Br(B^+\to X(3872)K^+)=(7.88^{+4.87}_{-3.76})\times10^{-4}$. Obviously, this result is much larger than the present experimental upper
limits given by Belle \cite{belle} and BaBar \cite{babar} at $90\%$ C.L. ,
\be
Br(B^+\to X(3872)K^+)&<&2.6\times10^{-4} \;\;\;(\text{Belle}) , \\
Br(B^+\to X(3872)K^+)&<&3.2\times10^{-4}  \;\;\;(\text{BaBar}).
\en

Here we would like to systematically study the decays $B_{c,u,d}\to X(3872)P$ in the perturbtive QCD
approach, where $P$ represents a light pseudoscalar meson K and $\pi$. The layout of this paper is as follows, we calculate analytically the
amplitudes for the studied
decays $B_{c,u,d}\to X(3872)P$ in Section II. The numerical results and discussions are given in Section III, where we will compare
our results with other theoretical predictions and the data. The conclusions are presented in the final part.

\section{the amplitudes of $B_{c,u,d}\to X(3872)P$ decays }
As we know, the PQCD factorization approach has been used to calculate many two-body charmed B meson
decays, and obtained consistent results with experiments. So we wil use this approach to study the decays
$B_{c,u,d}\to X(3872)P$. First, the effective Hamiltonian for the decays $B^+_c\to X(3872)\pi^+ (K^+)$ can be written as \cite{buchalla}
\be
\emph{H}_{eff}=\frac{G_F}{\sqrt2}V^*_{cb}V_{uq}\left[C_1(\mu)O_1(\mu)+C_2(\mu)O_2(\mu)\right]+H.c.,
\en
where $C_{1,2}(\mu)$ are the Wilson coefficients at the renormalization scale $\mu$,
$q=d$ ($q=s$) for $B^+_c\to X(3872)\pi^+$ ($B^+_c\to X(3872)K^+$) decay, and $O_{1,2}$ are the four fermion
operators,
\be
O_1=\bar d_{\alpha}\gamma_{\mu}(1-\gamma_5)u_{\beta}\otimes\bar c_{\beta}\gamma_{\mu}(1-\gamma_5)b_{\alpha},\;\;
O_2=\bar d_{\alpha}\gamma_{\mu}(1-\gamma_5)u_{\alpha}\otimes\bar c_{\beta}\gamma_{\mu}(1-\gamma_5)b_{\beta}.\;
\en
Here we take the decay $B^+_c\to X(3872)\pi^+$ as an example to analysis and its Feymman diagrams are given in Fig.1,
\begin{figure}[t]
\vspace{-4cm} \centerline{\epsfxsize=18 cm \epsffile{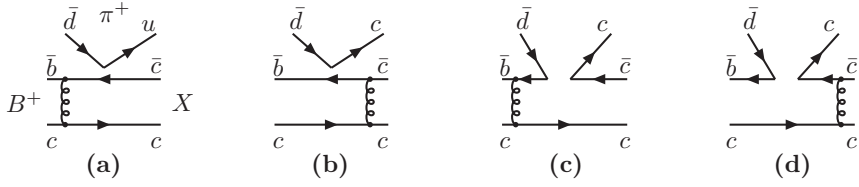}}
\vspace{-19.5cm} \caption{ Feynman diagrams contributing to the
decay $B^+_c\to X(3872)\pi^+$ at leading order.  }
 \label{fig1}
\end{figure}
where only the
factorizable and non-factorizable emission diagrams are needed considered at the leading order\footnote{From now on,
we will use $X$ to denote $X(3872)$ for simply in some places.}.
The amplitude for the factorizable emission diagrams Fig.1(a)
and Fig.1(b) can be written as
\be F^{LL}_{B_c\to X}&=&2\sqrt{\frac{2}{3}}\pi C_F
m^4_{B_c}f_{\pi}f_{B_c}\sqrt{1-r_X^2}\int_0^1 dx_2 \int_0^\infty b_1 db_1 b_2 db_2 exp(-\frac{\omega^2_{B_c}b^2_1}{2})\non&&
\left\{\left[
\Psi^L(x_2-2r_b)+\Psi^t(r_b-2x_2)\right]E_e(t_a)h(\alpha,\beta_a,b_1,b_2)S_t(x_2)\right.\non
&&\left.
-\Psi^L(r_c+r_X^2(x_1-1))E_e(t_b)h(\alpha,\beta_b,b_2,b_1)S_t(x_1)
\right\}, \en where the
color factor $C_F=4/3, f_{\pi(X)}$ is the decay constant for $\pi(X)$ meson, the mass ratio $r_{X(b,c)}=m_X(m_b,m_c)/m_B$,
and $\Psi^{L,t}$ are the
distribution amplitudes for the $X(3872)$ given in Sec. III. The evolution factors $E_e(t_{a,b})$ evolving the Sudakov exponent, the hard
functions $h$ and the jet function $S_t(x)$ can be found in
Ref. \cite{zhour0,zhang}. The amplitude for the non-factorizable spectator
diagrams Fig.1(c) and Fig.1(d) is given as
\be M^{LL}_{B_c\to X}&=&\frac{8}{3}\pi C_F m^4_{B_c}f_{B_c}\sqrt{1-r_X^2} \int_0^1  dx_2 \int_0^\infty
b_1 db_1 b_3 db_3 \phi^A_\pi(x_3)\non &&\left\{[\Psi^L(x_2)(x_3-x_1)(1-r_X^2)+r_X\Psi^t(x_2)(1-x_1-x_2)]E_c(t_c)h(\beta_c,\alpha,b_3,b_1)
\right.\non &&\left.
+[\Psi^L(x_2)(r_X^2(x_2-x_3)+2x_1+x_2+x_3-2)+r_X\Psi^t(x_2)(1-x_1-x_2)]\right.\non &&\left.
E_d(t_d)h(\beta_d,\alpha,b_3,b_1)\right\},
\label{cd1} \en
where $\alpha, \beta_{a,b,c,d}$ in the hard function and the hard scales $t_{a,b,c,d}$ are given in Appendix.

Second, the effective Hamiltonian for the decays $B_{u,d}\to X(3872)\pi(K)$
is written as
\be
\emph{H}_{eff}=\frac{G_F}{\sqrt2}\left[V^*_{cb}V_{cq}(C_1(\mu)O^c_1(\mu)+C_2(\mu)O^c_2(\mu))-
V^{*}_{tb}V_{tq}\sum^{10}_{i=3}C_i(\mu)O_i(\mu)\right],
\en
where $V^*_{c(t)b}V_{c(t)q}$ is the product of the Cabibbo-Kobayashi-Maskawa (CKM) matrix elements, $q=d$ or $s$. The local four-quark operators $O_i(\mu)$ and the corresponding QCD-corrected Wilson coefficients $C_i(\mu)$ can
be found in Ref. \cite{buchalla}. Here we take $B^+\to X(3872)\pi^+$ as an example to analysis and its Feynmman diagrams are given in Fig.2.
\begin{figure}[t]
\vspace{-4cm} \centerline{\epsfxsize=18 cm \epsffile{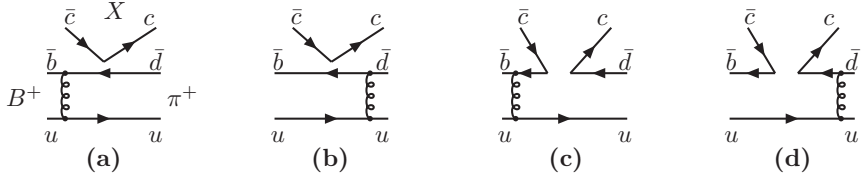}}
\vspace{-19.5cm} \caption{Feynman diagrams contributing to the decay
$B^+\to X(3872)\pi^+$ at leading order.}
 \label{Figure1}
\end{figure}
The amplitudes for the factorizable and the nonfactorizable emission diagrams from $(V-A)\otimes(V-A)$ operators are denoted as
$F^{LL}_{B\to \pi}$ and $M^{LL}_{B\to\pi}$, respectively. Their analytic expressions are given as
\be
F^{LL}_{B\to \pi}&=&\frac{8\pi C_F
m^4_Bf_{X}}{\sqrt{1-r^2_X}}\int_0^1 dx_1 dx_3 \int_0^\infty b_1 db_1 b_3 db_3\phi_B(x_1,b_1)
\left\{\left[
\left(r^2_X-1)\phi^A_\pi(x_3)((r^2_X-1)\right.\right.\right.\non&&\left.\left.\left. \times x_3-1\right) +(r^2_X-1)\phi^P_\pi(x_3)r_\pi(2x_3-1)+\phi^T_\pi(x_3)r_\pi(2x_3(r_X^2-1)+1+r^2_X)\right]
\right.\non
&&\left.\times E_e(t'_a)h(\alpha',\beta'_a,b_1,b_3)S_t(x_3)
-2r_\pi(1-r_X^2)\phi^P_\pi(x_3)\right.\non&&\left.\times E_e(t'_b)h(\alpha',\beta'_b,b_3,b_1)S_t(x_1)
\right\}, \en
\be M^{LL}_{B\to\pi}&=&\frac{32}{\sqrt6}\frac{1}{\sqrt{(1-r_X^2)}}\pi C_F m^4_B \int_0^1 dx_1 dx_2 \int_0^\infty
b_1 db_1 b_2 db_2\phi_B(x_1,b_1)\non &&\left\{\left[\Psi^L(x_2)(\phi^A_\pi(x_3)(r_X^2-1)+2\phi^T_\pi(x_3)r_\pi)(r_X^2(x_1+x_3-2x_2)+x_1-x_3)\right.\right.\non &&\left.\left.
+4r_\pi r_Xr_c\phi^T_\pi(x_3)\Psi^t(x_2)\right]\right\}E_c(t'_c)h(\beta'_c,\alpha',b_2,b_1).
\label{cd1} \en
Except the upper two $(V-A)\otimes(V-A)$ type amplitudes, there also exist facorizable and nonfacorizable emission contributions from $(V-A)\otimes(V+A)$
and $(S-P)\otimes(S+P)$ operators, which are expressed as $F^{LR}_{B\to\pi}$ and $M^{SP}_{B\to\pi}$, respectively,
\be
F^{LR}_{B\to\pi}&=&-F^{LL}_{B\to \pi},\\
 M^{SP}_{B\to\pi}&=&-\frac{32}{\sqrt6}\frac{1}{\sqrt{(1-r_X^2)}}\pi C_F m^4_B \int_0^1 dx_1 dx_2 \int_0^\infty
b_1 db_1 b_2 db_2\phi_B(x_1,b_1)\non &&\left\{\left[\Psi^L(x_2)(\phi^A_\pi(x_3)(r_X^2-1)+2\phi^T_\pi(x_3)r_\pi)(r_X^2(x_1+x_3-2x_2)+x_1-x_3)\right.\right.\non &&\left.\left.
-4r_\pi r_Xr_c\phi^T_\pi(x_3)\Psi^t(x_2)\right]\right\}E_c(t'_c)h(\beta'_c,\alpha',b_2,b_1),
\label{cd1} \en
where $\alpha',\beta'_{a,b,c}$ in the upper hard functions and the hard scales $t'_{a,b,c}$ are defined in Appendix.

By combining the amplitudes from the different
Feynman diagrams, the total decay amplitudes for the considered decays are given as
\be
\mathcal{A}(B_c\to X(3872)P)&=&V^*_{cb}V_{uq}\left[a_1F^{LL}_{B_c\to X}+C_1M^{LL}_{B_c\to X}\right],\\
\mathcal{A}(B_{u,d}\to X(3872)P)&=&V^*_{cb}V_{cq}\left[a_2F^{LL}_{B\to P}+C_2M^{LL}_{B\to P}\right]-V^*_{tb}V_{tq}\left[(a_3+a_9-a_5-a_7)F^{LL}_{B\to P}\right.\non &&\left.+(C_4+C_{10})M^{LL}_{B\to P}
+(C_6+C_8)M^{SP}_{B\to P}\right],\label{totalam}
\en
where the combinations of the Wilson
coefficients $a_1=C_1/3+C_2, a_2=C_1+C_2/3, a_i=C_i+C_{i+1}/3$ with $i=3,5,7,9$, $q=d$ and $q=s$ are corresponding to the decays induced by the $b\to d$ and $b\to s$
transitions, respectively.

\section{Numerical results and discussions} \label{numer}
We use the following input parameters for the numerical calculations
\cite{pdg20,WSL,hsiao} \be
f_{B_c}&=&0.398^{+0.054}_{-0.055} GeV,  f_{B}=0.19 GeV, f_{X}=0.234\pm0.052 GeV,\\
M_{B_{c}}&=&6.275 GeV, M_B=5.279 GeV, M_{X}=3.87169  GeV,\\
\tau_{B_c}&=&0.510\times 10^{-12} s, \tau_B^\pm=1.638\times 10^{-12} s,\tau_{B^0}=1.519\times 10^{-12} s. \en
For the CKM matrix elements, we adopt
the Wolfenstein parametrization and the updated values $A=0.814,
\lambda=0.22537, \bar\rho=0.117\pm0.021$ and
$\bar\eta=0.353\pm0.013$ \cite{pdg20}.
With the total amplitude, one can write the decay width as
\be
\Gamma(B\to X(3872)P)=\frac{G^2_F}{32\pi m_B}(1-r^2_{X})|\mathcal{A}(B\to X(3872)P)|^2.
\en
The wave functions of $B, \pi$ and $K$ have been well defined in many works, while those of the $B_c$ and $X(3872)$
exit many uncertainties. For the meson $B_c$, we use its wave function in the nonrelativistic limit \cite{liux1}
\be
\Phi_{B_c}(x)=\frac{if_{B_c}}{4N_C}\left[(\psl_{B_{c}} +M_{B_{c}})\gamma_5\delta(x-r_c)\right] \exp(-\frac{b^2\omega^2_{B_c}}{2}),
\en
where the last exponent term shows the $k_T$ dependence. For the $X(3872)$, the light cone distribution amplitude is taken the similar
formular with that of the $\chi_{c1}$ meson \cite{liux,chen}
\be
\langle X(3872)(p,\epsilon_L)|\bar c_\alpha(z)c_{\beta}(0)|0\rangle&=&\frac{1}{\sqrt{2N_c}}\int dxe^{ix p\cdot z}
\left\{m_X[\gamma_5\esl]_{\beta\alpha}\phi^L_X(x)+[\gamma_5\psl]_{\beta\alpha}\phi^t_X(x)\right\}.\;\;\;\;\;\;
\en
Here only the longitudinal polarization contributes to the considered decays, and $\phi^{L,t}_X(x)$ are given as
\be
\phi^{L}_X(x)&=&24.68\frac{f_X}{2\sqrt{2N_c}}x(1-x)\left\{\frac{x(1-x)(1-2x)^2\left[1-4x(1-x)\right]}{[1-3.47x(1-x)]^3} \right\}^{0.7},\\
\phi^{t}_X(x)&=&13.53\frac{f_X}{2\sqrt{2N_c}}(1-2x)^2\left\{\frac{x(1-x)(1-2x)^2\left[1-4x(1-x)\right]}{[1-3.47x(1-x)]^3} \right\}^{0.7}.
\en

Using the input parameters and the wave functions as specified
in this section, we give the branching ratios of the decays $B^+_c\to X(3872)\pi^{+}(K^+)$ as follows
\be
Br(B^+_c\to X(3872)\pi^{+})&=&(2.7^{+1.4+0.9+0.7+0.2}_{-1.0-0.6-0.5-0.1})\times10^{-4},\\
Br(B^+_c\to X(3872)K^{+})&=&(2.5^{+1.3+0.8+0.6+0.2}_{-1.0-0.6-0.4-0.1})\times10^{-5},
\en
where the first error comes from the decay constant of the meson $X(3872)$, $f_{X}=0.234\pm0.052$ GeV,
the second and third uncertainties is caused by the shape parameter $\omega_{B_c}=0.6\pm0.1$ GeV and the decay constant $f_{B_c}=0.398^{+0.054}_{-0.055}$ GeV
, respectively, the last error is from the hard scale-dependent uncertainty, varied from $0.8t$ to $1.2t$. One can find that the branching ratios are
sensitive to the decay constant $f_{X}$. This is because that the dominate contributions for these two channels are from the factorization emission
amplitudes, which are proportional to $f_{X}$. $Br(B^+_c\to X(3872)\pi^{+})$ is about one order larger than $Br(B^-_c\to X(3872)K^{-})$, which is mainly
induced by the different from CKM elements $V_{ud}=1-\lambda^2/2$ and $V_{us}=\lambda$. From Table \ref{tab1}, it is shown that our predictions are
consistent with the results given in the
covariant light-front quark model within errors \cite{WSL}, but much larger than those calculated by the generalized factorization approach \cite{hsiao}.
\begin{table}
\caption{Our predictions for the branching ratios of the decays $B^+_{c}\to X(3872)\pi^{+}(K^{+})$, together with the results from
the covariant light-front approach (CLFA) \cite{WSL} and the generalized factorization
approach (GFA) \cite{hsiao}. }
\begin{center}
\begin{tabular}{c|c|c|c}
\hline\hline mode& This work &CLFA\cite{WSL}& GFA\cite{hsiao} \\
\hline
$B^+_{c}\to X(3872)\pi^{+}(\times10^{-4})$&$2.7^{+1.4+0.9+0.7+0.2}_{-1.0-0.6-0.5-0.1}$&$1.7^{+0.7+0.1+0.4}_{-0.6-0.2-0.4}$&$0.60^{+0.22+0.14}_{-0.18-0.07}$ \\
$B^+_{c}\to X(3872)K^{+}(\times10^{-5})$&$2.5^{+1.3+0.8+0.6+0.2}_{-1.0-0.6-0.4-0.1}$&$1.3^{+0.5+0.1+0.3}_{-0.5-0.2-0.3}$&$0.47^{+0.17+0.11}_{-0.14-0.05}$ \\
\hline\hline
\end{tabular}\label{tab1}
\end{center}
\end{table}

Similarly, the branching ratios of the decays $B\to X(3872)P$ are calculated as follows
\be
Br(B^+\to X(3872)K^{+})&=&(3.8^{+0.9+0.6+0.3}_{-0.8-0.5-0.2})\times10^{-4},\\
Br(B^0 \to X(3872)K^{0})&=&(3.5^{+0.7+0.5+0.3}_{-0.6-0.4-0.2})\times10^{-4},\\
Br(B^+\to X(3872)\pi^{+})&=&(9.3^{+1.5+0.9+0.5}_{-1.3-0.8-0.4})\times10^{-6},\\
Br( B^0\to X(3872)\pi^{0})&=&(4.3^{+0.7+0.5+0.3}_{-0.6-0.4-0.3})\times10^{-6},
\en
where the first uncertainty comes from the shape parameter $\omega_{B}=0.4\pm0.04$ GeV in $B$ meson wave function, the second error is
from the decay constant $f_{X}=0.234\pm0.052$ GeV of the $X(3872)$, and the third one arises from the choice of the hard scales,
which vary from $0.8t$ to $1.2t$. From the results, one can find that the branching ratios for the decays $B^+\to X(3872)K^{+}$ and $ B^0 \to X(3872)K^{0}$ are close to each other, since they differ only
in the lifetimes between $B^+$ and $ B^0$ in our formalism. Our prediction for the decay $B^-\to X(3872)K^{-}$ is less than the previous
PQCD calculation result $(7.88^{+4.87}_{-3.76})\times10^{-4}$ \cite{liux}.
While it is still slightly larger than the upper limits $2.6\times10^{-4}$ given by Belle \cite{belle} and $3.2\times10^{-4}$ given
by BaBar \cite{babar}.If the present experimental upper limits are confident, a pure charmonium assignment for
the $X(3872)$ is maybe not suitable under the PQCD approach.
 We expect that the branching ratios for the decays $B^{0,+}\to X(3872)K^{0,+}$ can be precisely measured
at the present LHCb and SuperKEKB experiments, which is very helpful to probe the inner structure of the $X(3872)$.

On the other hand, it is noticed that the $X(3872)$ is renamed as $\chi_{1c}(3872)$ by the present Particle Data Group (PDG) 2020 \cite{pdg20}, which seems to assume it as the radial
excited state of $\chi_{1c}(1P)$. As we know, the $\chi_{1c}(1P)$ is another P-wave charmonium state with the same quantum numbers
$J^{(PC)}=1^{++}$ but a slightly lighter mass $3.511$ GeV. In this case, they should have similar characters in the B meson decays. For example,
the branching ratio for the decay $B^{+}\to \chi(1P) K^+$ is measured as $(4.85\pm0.33)\times10^{-4}$ \cite{pdg20}, which is consistent with the result
predicted by PQCD approach $(4.4^{+1.9}_{-1.6})\times10^{-4}$ \cite{zhour}. The corresponding decay $B^{+}\to X(3872) K^+$ should have similar but
slightly smaller branching ratio.
The comparisons of the branching ratios between the decays $B\to X(3872)\pi(K)$
and $B\to \chi_{1c}(1P)\pi(K)$ can be found in Table \ref{tab2}, where the theoretical predictions for the $Br(B\to \chi_{1c}(1P)\pi(K))$
are taken from another PQCD calculations \cite{zhour}.  From Table \ref{tab2}, we can know that the calculations for the decays
$B\to X(3872)P$ by using the PQCD approach are under control and credible. So we suggest that the experimental colleagues can measure these decays at LHCb
and Belle II, which is helpful to discriminate the inner structures of the $X(3872)$ from different assumptions.
\begin{table}
\caption{Comparisons between $Br(B\to X(3872)\pi(K))$ (this work) and $Br(B\to \chi_{1c}(1P)\pi(K))$ \cite{zhour}
calculated in PQCD approach, and the data are taken from the Particle Data Group (PDG) 2020 \cite{pdg20}.   }
\begin{center}
\begin{tabular}{c|c|c|c|c}
\hline\hline mode$(\times10^{-4})$&$B^+\to X(3872)K^{+}$ &$B^+\to \chi_1(1P)K^{+}$ &$B^0\to X(3872)K^{0}$&$B^0\to \chi_1(1P)K^{0}$ \\
\hline
PQCD&$3.8^{+0.9+0.6+0.3}_{-0.8-0.5-0.2}$&$4.4^{+1.9}_{-1.6}$&$3.5^{+0.7+0.5+0.3}_{-0.6-0.4-0.2}$&$4.1^{+1.8}_{-1.6}$ \\
Exp.&--&                                $4.85\pm0.33$        &--                       &$3.95\pm0.27$\\
\hline
mode($\times10^{-5})$&$B^+\to X(3872)\pi^{+}$ &$B^+\to \chi_1(1P)\pi^{+}$ &$B^0\to X(3872)\pi^{0}$&$B^0\to \chi_1(1P)\pi^{0}$ \\
\hline
PQCD&$0.93^{+0.15+0.09+0.05}_{-0.13-0.08-0.04}$&$1.7\pm0.6$&$0.43^{+0.07+0.05+0.03}_{-0.06-0.04-0.03}$&$0.8\pm0.3$ \\
Exp.&--&                               $2.2\pm0.5$        &--                       &$1.12\pm0.28$\\
\hline\hline
\end{tabular}\label{tab2}
\end{center}
\end{table}

In Table \ref{tab3}, we compare our predictions with the results calculated in the generalized factorization approach \cite{hsiao}.
It is interesting that the branching ratios for the decays $B\to X(3872)\pi(K)$ calculated in these two different approaches are consistent
with each other within errors. One can find that
$Br(B^+\to X(3872)\pi^{+})\simeq2 Br(B^0\to X(3872)\pi^{0})$, which is supported by
the isospin symmetry.

\begin{table}
\caption{Our predictions for the branching ratios of the decays $B\to X(3872)\pi(K)$, together with the results from
the generalized factorization approach (GFA) \cite{hsiao}. }
\begin{center}
\begin{tabular}{c|c|c}
\hline\hline mode& This work &GFA \cite{hsiao} \\
\hline
$B^+\to X(3872)K^{+}(\times10^{-4})$&$3.8^{+0.9+0.6+0.3}_{-0.8-0.5-0.2}$&$2.3^{+1.1}_{-0.9}\pm0.1$ \\
$B^0\to X(3872)K^{0}(\times10^{-4})$&$3.5^{+0.7+0.5+0.3}_{-0.6-0.4-0.2}$&$2.1^{+1.0}_{-0.8}\pm0.1$ \\
$B^+\to X(3872)\pi^{+}(\times10^{-6})$  &$9.3^{+1.5+0.9+0.5}_{-1.3-0.8-0.4}$&$11.5^{+5.7}_{-4.5}\pm0.3$ \\
$B^0\to X(3872)\pi^{0}(\times10^{-6})$&$4.3^{+0.7+0.5+0.3}_{-0.6-0.4-0.3}$&$5.3^{+2.6}_{-2.1}\pm0.2$ \\
\hline\hline
\end{tabular}\label{tab3}
\end{center}
\end{table}

In the following we will discuss the CP-violating asymmetries of the decays $B\to X(3872)P$.
As we know, the CP
asymmetry arises from the interference between the tree and penguin amplitudes, while for the decays
$B^+_c\to X(3872)\pi^+(K^+)$, there are no contributions from the penguin amplitudes, so the corresponding direct CP violation is zero.
For the charged decays $B^+\to X(3872)\pi^{+}(K^+)$, we need only consider the direct CP violation $A^{dir}_{CP}$, which is defined as
\be
A^{dir}_{CP}=\frac{|\mathcal{\bar A}|^2-|\mathcal{A}|^2}{|\mathcal{ \bar A}|^2+|\mathcal{A}|^2},
\en
where $\mathcal{ \bar A}$ is the CP-conjugate amplitude of $\mathcal{A}$. As to the neutral B meson decays,
there exists another type CP violation named as time-dependent CP asymmetry, which is induced by the interference
between the direct decay and the decay via oscillation, need to be considered. Then the time-dependent CP violation
can be defined as
\be
A(t)_{CP}=A_fcos(\Delta mt)+S_fsin(\Delta mt),
\en
where $\Delta m>0$ is the mass difference of the two neutral B meson mass eigenstates, and the direct CP asymmetry $A_f$ and
the mixing-induced CP asymmetry are expressed as
\be
A_f=\frac{|\lambda_f|^2-1}{|\lambda_f|^2+1}, \;\;S_f=\frac{2Im(\lambda_f)}{|\lambda_f|^2+1}, \label{ASf}
\en
with $\lambda_f=\eta_f e^{-2i\beta}\frac{\mathcal{\bar A}}{\mathcal{A}}$. $\eta_f$ is $1(-1)$ for a CP-even (CP-odd) final state $f$, $\beta$
is the CKM angle \cite{pdg20}.
Since the charged decay channel and the corresponding neutral mode are the same with each other except the lifetimes and isospin factor in the
amplitudes, they have the same direct CP asymmetries. So we need only consider the neutral decays, whose direct CP asymmetries are calculated as
\be
A_{X(3872)K_S}&=&(1.2^{+0.0+0.0+0.2}_{-0.0-0.0-0.3})\times10^{-3},\\
A_{X(3872)\pi^0}&=&(2.7^{+0.1+0.0+0.4}_{-0.2-0.0-0.4})\times10^{-2},
\en
where the errors are induced by the same sources as those for the branching ratios, but the difference is that the direct CP violations are
less sensitive
to the nonperturbative parameters within their uncertainties except the hard scale $t$. Compared with the tree contributions, the penguin amplitudes
are loop suppressed and lower by $1\sim 2$ orders of magnitude. At the same time, the product of the CKM matrix elements associated with
the tree amplitudes is about 4 times than that for penguin ones. So the direct CP violations, which arise from the interference
between the tree and penguin contributions, are very small.
Since the final state $X(3872)K^0$ and its CP conjugate are flavor-specific, so we should use the CP-odd eigenstate $X(3872)K_S$
to analyze the mixing-induced CP violations. The results  for the mixing-induced CP violations are given as
\be
S_{X(3872)K_S}&=&(70.3^{+0.0+0.0+0.9}_{-0.0-0.0-1.2})\%,\\
S_{X(3872)\pi}&=&(-60.8^{+0.0+0.0+1.5}_{-0.0-0.0-1.4})\%,
\en where the errors are similar with those listed in the direct CP
violations, and not sensitive to the nonperturbative parameters
given in the wave functions. One can find that $S_{X(3872)K_S}$ is consistent well
with the current world average value $\sin2\beta=0.699\pm0.017$
\cite{amhis}, which are obtained from $B^0$ to charmonium and $K^0$
decays. So we can check the nature of $X(3872)$ by extracting the
CKM phase $\beta$ from the future experimental data for the decay $B^0\to
X(3872)K_S$. The mixing-induced CP asymmetry for the decay
$B^0\to X(3872)\pi^0$ is some different from the world average value of $\sin2\beta$, which is because that the imagine parts of the total amplitudes
for this channel and its conjugate one exist larger different. Our results can be tested
by the future experiments.
\section{Summary}
We studied the decays $B_{c,u,d}\to X(3872)\pi(K)$ in the PQCD approach, where the puzzling resonance $X(3872)$
is involved. Assuming the $X(3872)$ as a $1^{++}$ charmonium state, we calculated the branching ratios and CP
asymmetries for the considered decays. Through comparing our predictions with other theoretical results and the available
experimental data, we found the following results

(1) The branching ratios of the decays $B^-_c\to X(3872)\pi^-$ and $B^-_c\to X(3872) K^-$ can reach the order of $10^{-4}$
and $10^{-5}$, respectively, which are consistent with the covariant light-front approach within errors, but larger than those given by the generalized
factorization approach. These results can be discriminated at the present running LHCb and Belle II experiments.

(2) Our predictions for the branching ratio of the decays $B\to X(3872) K$ and $B\to X(3872)\pi$ are consistent with the results given
by the covariant light-front approach. $Br(B\to X(3872) K)$ can reach the order of $10^{-4}$, which is much larger than that of
the decay $B\to X(3872) \pi$ induced by the $b\to d$ transition. On the experimental side, it is helpful to probe the inner structure
of the $X(3872)$
by measuring the branching ratios and testing the $SU(3)$ and
isospin symmetries for these considered decays.

(3) The direct CP violations for the decays $B\to X(3872)\pi(K)$ are very small, only $10^{-3}\sim 10^{-2}$. The mixing-induced CP violation
for the decay $B\to X(3872)K_S$ agrees well with current world average value $(69.9\pm1.7)\%$. But it is different
for the value of $S_{X(3872)\pi^0}$, which is because that the imagine parts of the total amplitudes
for the decay $B\to X(3872)\pi^0$ and its conjugate one exist larger different.
\section*{Acknowledgment}
This work is partly supported by the National Natural Science
Foundation of China under Grant No. 11347030, by the Program of
Science and Technology Innovation Talents in Universities of Henan
Province 14HASTIT037.
\appendix
\section{}
\be
\alpha&=&(x_2+x_1-1)(r_X^2(1-x_2)-x_1)m^2_{B_c}, \\
\beta_a&=&(r_b^2-x_2(1-r^2_X(1-x_2))m^2_{B_c}, \\
\beta_b&=&(r_c^2-(1-x_1)(r^2_X-x_1))m^2_{B_c}, \\
\beta_c&=&-(1-x_1-x_2)(r_X^2(1-x_2-x_3)+x_3-x_1)m^2_{B_c},\\
\beta_d&=&-(1-x_1-x_2)(r_X^2(x_3-x_2)+1-x_3-x_1)m^2_{B_c},\\
\alpha'&=&x_1x_3(1-r_X^2)m^2_{B},\\
\beta'_a&=&x_3(1-r_X^2)m^2_{B},\\
\beta'_b&=&x_1(1-r_X^2)m^2_{B},\\
\beta'_c&=&(r_c^2+(x_1-x_2)(x_3+r_X^2(x_2-x_3)))m^2_B,\\
t_{a(b)}&=&\max(\sqrt{|\alpha|},\sqrt{|\beta_{a(b)}|},1/b_1,1/b_2),\\
t_{c(d)}&=&\max(\sqrt{|\alpha|},\sqrt{|\beta_{c(d)}|},1/b_1,1/b_3),\\
t^\prime_{a(b)}&=&\max(\sqrt{|\alpha'|},\sqrt{|\beta'_{a(b)}|},1/b_1,1/b_3),\\
t'_c&=&\max(\sqrt{|\alpha'|},\sqrt{|\beta'_c|},1/b_1,1/b_2).
\en

\end{document}